\documentclass{jpsj2}
\title{Model for coiling and meandering instability  of viscous threads}

\author{Shin-ichiro \textsc{Nagahiro}$^{1}$\thanks{Multiple authors and affiliations correspond using arabic numerals each other.} and Yoshinori \textsc{Hayakawa}$^{2}$}

\inst{$^{1}$ Department of Mechanical Engineering, Miyagi National College of Technology, Miyagi 981-1239, Japan\\
$^{2}$Department of Physics, Tohoku University, Aoba-ku, Sendai, 980-8578, Japan}

\abst{A numerical model is presented to describe both the transient and steady-state dynamics of viscous threads falling onto a plane. The steady-state coiling frequency $\Omega$ is calculated as a function of fall height $H$. In the case of weak gravity, $\Omega\propto H^{-1}$ and $\Omega\propto H$ are obtained for lower and higher fall heights respectively. When the effect of gravity is significant, the relation $\Omega \propto H^2$ is observed. These results agree with the scaling laws previously predicted. The critical Reynolds number for coil-uncoil transition is discussed. When the gravity is weak, the transition occurs with hysteresis effects. 
If the plane moves horizontally at a constant speed, a variety of meandering oscillation modes can be observed experimentally. The present model also can describe this phenomenon. The numerically obtained state diagram for the meandering modes qualitatively agrees with the results of experiment.}

\kword{buckling instability, liquid rope coiling, free surface flow, numerical model}

\begin{document}
\maketitle

\section{Introduction} 
The instability of viscous fluid generally occurs when the Reynolds number exceeds a critical value, because, in the low-Reynolds-number regime, the eigenmodes with short wavelength are impeded. However, when the fluid surface can move freely and deform greatly, this is not the case. For example, an axisymmetric jet emitted from a nozzle onto a horizontal plane shows buckling instability {\it below} a critical Reynolds number.
The ``rope" of fluid loops near the plane and forms a coil owing to the buckling. This phenomenon is thus called "fluid rope coiling" and has been studied for several decades in the laboratory \cite{barnes1, barnes2, cruick1, cruick2, yarin}. Barnes and Woodcock first performed an experimental study and observed that the coiling frequency increases proportionally to the fall height\cite{barnes1,barnes2}. A more comprehensive investigation was conducted by Cruickshank and Munson\cite{cruick1}, who found the existence of the critical Reynolds number above which bucking does not occur. They also found that the coiling frequency is not a monotonically increasing function of fall height but decreases for low fall height.
The theoretical treatment of fluids with freely moving surfaces is not easy, however, some researchers have succeeded in determining the critical fall height and frequency at the onset of coiling using linear stability analysis with some simplifications and assumptions \cite{cruick2, yarin}.

Phenomenologically, one can understand that steady coiling proceeds with a mechanical balance between the driving force of a steady flow and the internal viscous stress in the buckling portion of the rope. \cite{mahadevan} assumed that, in the high frequency limit, the inertial force $F_I$ is equal to the viscous force $F_V$ in magnitude. Representing the flow rate as $Q$, the radius of the rope in the coil as $a$, and the kinetic viscosity as $\nu$, this mechanical balance yields a scaling law for the coiling frequency,
\begin{equation}
\Omega_I\propto \left(\frac{Q^4}{\nu a^{10}}\right)^{1/3}\label{eq:inertial}
\end{equation}
which is called ``inertial coiling". 
Considering the gravitational force $F_G$ and the fluid injection force $F_P$, \cite{ribe_prsl} asserted that there are two more scaling laws. "Gravitational coiling" takes place under the condition $F_G\sim F_V$, and "Viscous coiling" when $F_P\sim F_V$. Each mechanical balance yields
\begin{align}
\Omega_G&\propto \left(\frac{gQ^3}{\nu a^8}\right)^{1/4},\label{eq:gravitational}\\
\Omega_V&\propto \frac{Q}{Ha^2},\label{eq:viscous}
\end{align}
where $g$ is the gravitational acceleration and $H$ is the fall height.
The existence of these three distinct coiling regimes were confirmed experimentally \cite{ribe_prl, ribe_pre}.

Ribe also derived the differential equations for a very thin rotating rope to predict the steady-state coiling frequency as a function of fall height, and demonstrated the three different coiling regimes. His analysis revealed that the steady solution is multivalued so that there might a discontinuity in the selected coiling frequency \cite{ribe_prsl, ribe_pf1}. 

The theories of the buckling and coiling instability have been restricted to the description of the steady-state coiling frequency or the onset of buckling with infinitesimal amplitudes. 
Furthermore, a recent experiment revealed that fluid rope falling onto a moving belt shows a rich variety of ``meandering" patterns \cite{webstar, ribe_pf2, morris}, but a theory that successfully explains the state diagram for various patterns has not yet been proposed.
Hence a numerical model that can describe the entire dynamics of viscous fluid rope is still needed.

In this study, we use our recently proposed numerical model \cite{nagahiro} to understand the transient dynamics of the coil-uncoil transition and steady-state coiling frequency. We also extend the model to describe the meandering instability of fluid rope on a moving surface.

This paper is organized as follows. In section \ref{sec:model}, partial differential equations that describe the dynamics of coiling are derived. Coil-uncoil transition and its hystereric effects are discussed in section \ref{sec:coil-uncoil}. The coiling frequency as a function of fall height is discussed in section \ref{sec:steady_freq}.  In section \ref{sec:meandering},  the present model is applied to the problem of the meandering instability of fluid rope falling onto a moving surface. In section \ref{sec:summary}, we summarize our results.

\section{Model}\label{sec:model}
\begin{figure}
  \centerline{\includegraphics[width= 3.5cm, keepaspectratio]{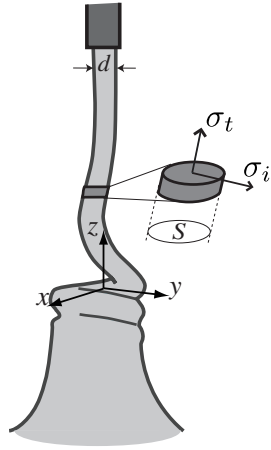}}
  \caption{A schematic view of fluid rope model.}\label{fig:model}
\end{figure}
We present a simple numerical model for a thread of viscous fluid falling onto a plane. Analogous equations were used by Chiu-webstar and Lister to illustrate the steady-state viscous catenary dragged by a horizontally moving belt\cite{webstar}.

The present model is restricted to the case of slight deformation: viscous stiffness for bending and twisting are neglected. We assume uniform flow within the rope and do not include the surface tension effect. Consider a viscous fluid ejected from an orifice at a sufficient height. We require that the upward growth speed of the coil be balanced by the downward slumping speed at the top of the coil, namely, the top of coiling portion does not move. Under this assumption, we fix the origin of the reference frame at the point where the fluid rope begins to coil (see Fig. \ref{fig:model}) and only consider the flow at $z>0$.

Because of the assumption of slight deformation, the unit tangential vector of the rope ${\boldsymbol t}=(t_x,t_y,t_z)$ should be almost parallel to the $z$ axis. We thus replace the derivative with respect to $\boldsymbol t$ by that of $z$. 
Let $S$ be the cross-sectional area parallel to the $xy$ plane, and $w$ the axial flow velocity. The conservation of the volume flux is written as 
\begin{equation}
\frac{\partial S}{\partial t}=-\frac{\partial}{\partial z}(Sw).\label{eq:consev_volume}
\end{equation}
Let ${\boldsymbol q}=(q_x, q_y)$ be the center of mass at a given height, and ${\boldsymbol u}=(u_x,u_y)$ be the velocity; these obey
\begin{equation}
\left(\frac{\partial}{\partial t}+w\frac{\partial}{\partial z}\right)q_i=u_i,
\label{eq:eom_q}
\end{equation}
where $i=x$ and $y$. We denote the axial stress acting on the cross section as $\sigma_t$, which would be the driving force of the oscillation, and the viscous shear stress as $\sigma_{i}$, which would be the resistance force to bending. 
The equation of motion for $w$ is 
\begin{equation}
\left(\frac{\partial}{\partial t}+w\frac{\partial}{\partial z}\right)w
=\frac{1}{\rho S}\frac{\partial}{\partial z}\left(S\sigma_{t}\right)-g,\label{eq:eom_w}
\end{equation}
where $\rho$ is the density of fluid. The $i$th component of the stress acting on a cross section is $\sigma_i+t_i\sigma_t$. Therefore, we obtain the equation of motion for $u_i$ as
\begin{equation}
\left(\frac{\partial}{\partial t}+w\frac{\partial}{\partial z}\right)u_i=
\frac{1}{\rho S}\frac{\partial}{\partial z}
\left\{S\left( \sigma_i+t_i \sigma_t\right)\right\}.
\label{eq:eom_u}
\end{equation}

The shear stress can be written as $\sigma_i = \eta\partial(u_i+wt_i)/\partial z$, with $\eta$ as the viscosity. We can easily find that the axial stress $\sigma_{t}$ does not explicitly include the fluid pressure $p$, as follows. Let $r$ be the radial coordinate for the local cross section, and $u_r$ the flow velocity in the direction $r$; the conservation of volume flux gives $2(\partial u_r/\partial r)=-\partial w/\partial z$. The radial stress can be written as 
$\sigma_{r}=-p+2\eta(\partial u_r/\partial r)=-p-\partial w/\partial z$. 
The radial stress must vanish at the free surface, thus we obtain $p=-\eta(\partial w/\partial z)$. The axial stress, therefore, can be written as \cite{trouton}
\begin{equation}
\sigma_{t}=-p-2\eta\frac{\partial w}{\partial z}=-3\eta\frac{\partial w}{\partial z}.
\label{eq:axial_stress}
\end{equation}
Using the expressions of $\sigma_t$ and $\sigma_i$, the dimensionless forms of Eqs. (\ref{eq:eom_w}) and (\ref{eq:eom_u}) become
\begin{eqnarray}
\left(\frac{\partial}{\partial t}+w\frac{\partial}{\partial z}\right)w&=&\frac{3}{S\hspace{0.5mm}{\rm Re}}\frac{\partial}{\partial z}\left(S\frac{\partial w}{\partial z}\right) -\frac{1}{\rm Fr},
 \label{eq:dimless_eom_w}\\
\left(\frac{\partial}{\partial t}+w\frac{\partial}{\partial z}\right)u_i&=&
\frac{1}{S\hspace{0.5mm}{\rm Re}}\frac{\partial}{\partial z}
\left\{S\left(\frac{\partial u_i}{\partial z}+4t_i\frac{\partial w}{\partial z}+w\frac{\partial t_i}{\partial z}\right)\right\} \label{eq:dimless_eom_u},\nonumber\\
\end{eqnarray}
where ${\rm Re}=d|w_{in}|/\nu$ is the Reynolds number and ${\rm Fr}=w_{in}^2/gd$ is the Froude number. Note that the vector $\boldsymbol t$ is determined from the derivative of $\boldsymbol q$ with respect to $z$ as ${\boldsymbol n} = \left({\partial q_x}/{\partial z} , {\partial q_y}/{\partial z}, 1\right)/C$ with $C=\sqrt{(\partial q_x/\partial z)^2+(\partial q_y/\partial z)^2+1}$.

Next, we discuss the boundary condition of the present model. At the neighborhood of the orifice, we neglect the relaxation of Poiseuille flow to plug flow. Thus, at the injection point $z=H/d$,
\begin{equation}
q_i(H/d)=0,~~u_i(H/d)=0,~~S(H/d)=\frac{\pi}{4},~~w(H/d)=-1.
\end{equation}
At $z=0$, we assume the free-end boundary condition
\begin{equation}
q_i'(0) = 0.~~ u_i'(0)=0, \label{eq:bc_bottom1}
\end{equation}
where the prime indicates a derivative with respect to $z$. To determine the boundary value of the axial flow velocity $w(0)$, we utilize a phenomenological parameter, the ``energy loss coefficient $(\equiv\alpha)$" proposed by Cruickshank\cite{cruick3}. They postulated that the rope starts to buckle at the height $z=\zeta$ where the viscous stress changes its sign. Considering energy, momentum and volume flux conservation across the buckling region $0<z<\zeta$, they derived $w(0)/w(\zeta)=(1-\alpha)/(1+\alpha)$.
The value of $\alpha$ is experimentally determined as $0.76$, which is fairly independent of viscosity, flow rate and orifice diameter. 
Therefore, we require the following time dependent boundary condition:
\begin{equation}
w(0) = \beta w(\zeta),~~~~\beta=0.14. \label{eq:bc_bottom2}
\end{equation}
Note that, because of Eq. (\ref{eq:axial_stress}), $w(\zeta)$ is the maximum axial velocity that can be determined by solving Eq. (\ref{eq:dimless_eom_w}). How the present model is affected by the value of $\beta$ is discussed in the next section.

We numerically solve the partial differential equations (\ref{eq:consev_volume}), (\ref{eq:eom_q}), (\ref{eq:dimless_eom_w}) and (\ref{eq:dimless_eom_u}) using the Euler scheme. The space interval $\Delta z=0.1$ is fixed and the time step is set as $\Delta t=0.25{\rm Re}\Delta z^3$. Control parameters are Reynolds number ${\rm Re}$, Froude number {\rm Fr}, and fall height $H/d$. 
The simulation starts with the initial conditions $w(z)=-1$, $s(z)=\pi/4$, $q_i(z)=0$ and $u_i(z)=0$. A small roughness with an amplitude of $0.01$ is given to $q_i(z)$ as an initial shape at $t=0$.

\begin{figure}
  \centerline{\includegraphics[width= 9cm, keepaspectratio]{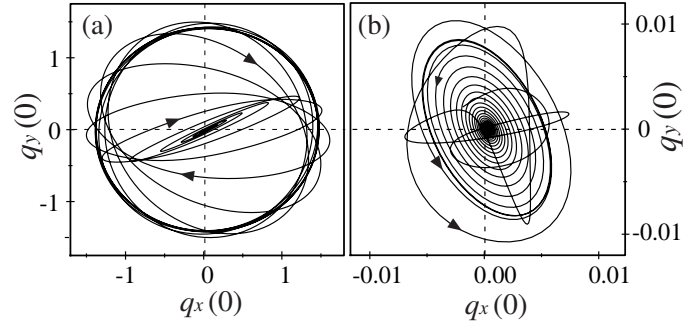}}
  \caption{Trajectories of the bottom of the model rope starting at $t=0$. 
(a) Trajectory when the circular coiling motion is stable in the steady state (Re=1.0, Fr=1.0, H/d=10.0). (b) Trajectory that converges to stable axial flow. Parameters are the same as in (a) but Reynolds number is slightly larger (Re=3.0).} \label{fig:2_orbits}
\end{figure}
\begin{figure}
  \centerline{\includegraphics[width= 8cm, keepaspectratio]{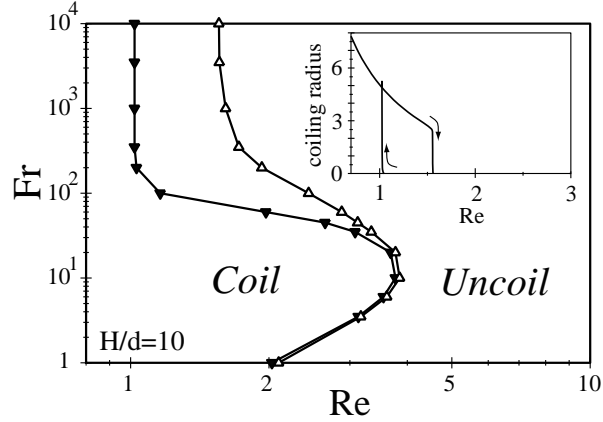}}
  \caption{Coiling regime in the Re-Fr plane for increasing (upward triangles) and decreasing (downward triangles) Reynolds number. The inset shows the radius of coiling for Fr $=10^4$ as a function of Re. } \label{fig:phase_FR}
\end{figure}
\section{Coil-uncoil transitions}\label{sec:coil-uncoil}
It is empirically known that fluid threads do not buckle if the Reynolds number of the flow is sufficiently large. Experimentally, the critical value of Reynolds number for coil-uncoil transition ranges from $0.7$ to $2.0$ and in average $1.2$, regardless of the value of Fr \cite{cruick1}. The present model correctly reproduces this well-known behavior. In Fig. \ref{fig:2_orbits}(a), we plot the trajectory of $\boldsymbol q$ in the plane $z=0$ for ${\rm Re}={\rm Fr}=1$ and $H/d=10$. The initial roughness of $\boldsymbol q(z)$ gradually increases and the trajectory converges to a circle with a radius of order unity. However, in the case of Re larger than the critical value $({\rm Re}=3)$, the amplitude of oscillation monotonically decreases, as shown in Fig. \ref{fig:2_orbits}(b), and the axial stagnation flow becomes stable. 

We investigate critical Reynolds number ${\rm Re}^*$, and find that the transition occurs with a hysteresis effect. In the simulation, we change the value of Re sufficiently slowly (slower than 0.1\% change per cycle of motion) after the motion of fluid reaches a steady state. A small perturbation of the amplitude, $10^{-6}d$, is continuously added for $q_i$ throughout the time evolution. In this way, we obtain the coiling radius as a function of increasing or decreasing Re, as shown in the inset of Fig. \ref{fig:phase_FR}. For increasing {\rm Re}, the radius sharply falls at ${\rm Re}=1.6~(\equiv {\rm Re}^*_h)$ and for decreasing Re, the radius rises at ${\rm Re}=1.0~(\equiv {\rm Re}^*_l)$. The axial stagnation flow is absolutely stable at ${\rm Re}>{\rm Re}^*_h$ and is absolutely unstable at ${\rm Re}<{\rm Re}^*_l$. In the bistable state (${\rm Re}^*_l<{\rm Re}<{\rm Re}^*_h$), the fluid thread starts to coil if the amplitude of perturbation exceeds unity in order. The difference ${\rm Re}^*_h-{\rm Re}^*_l$ reaches the maximum at ${\rm Fr}=10^2$ and disappears for ${\rm Fr}<10$. Although the hysteresis effect in the coil-uncoil transition has not been observed experimentally, such behavior can occur owing to the inertial effect (centrifugal force keeps the fluid thread rotating).

\begin{figure}
  \centerline{\includegraphics[width= 9cm, keepaspectratio]{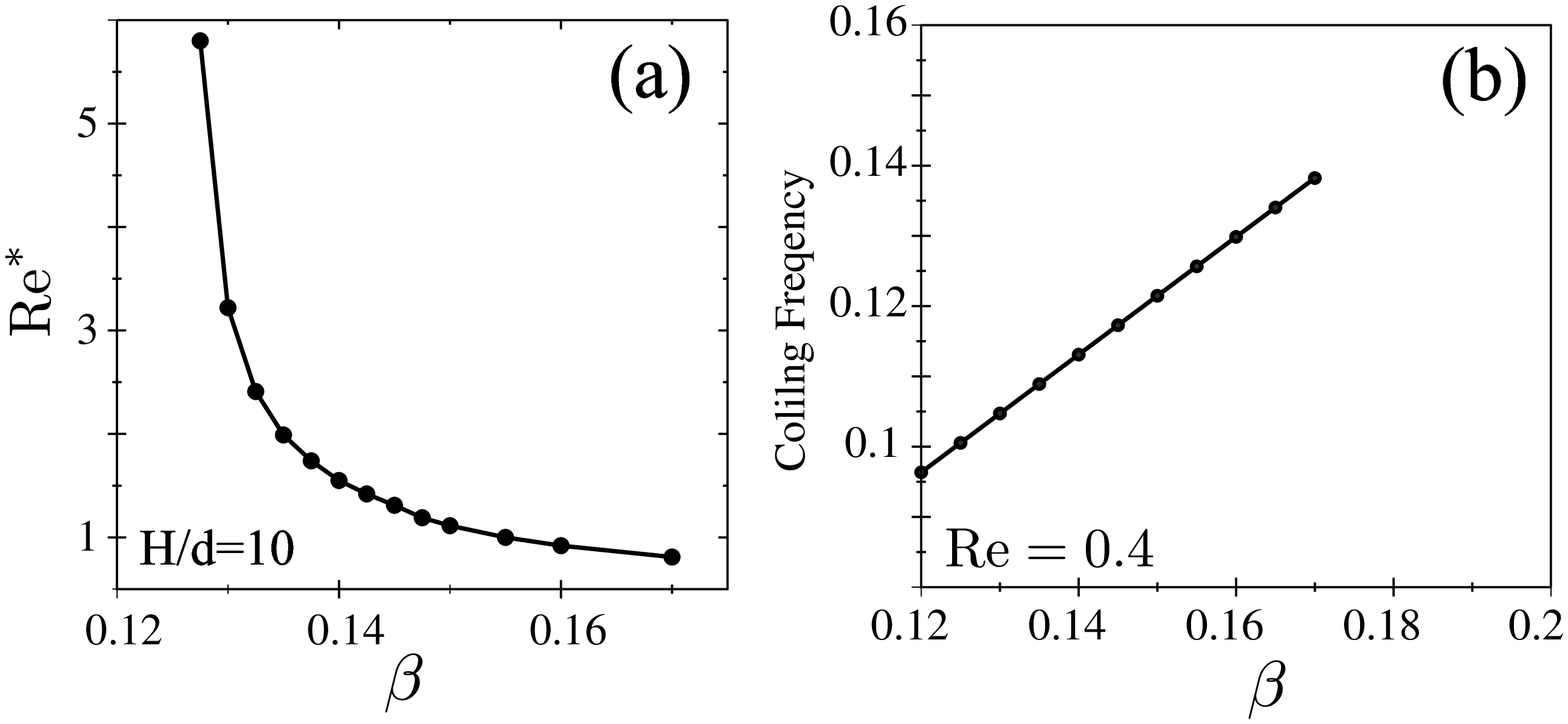}}
  \caption{(a) Critical Reynolds number ${\rm Re}_h^*$ versus $\beta$. (b) Steady-state coiling frequency as a function of $\beta$ for ${\rm Re}=0.4$. Other parameters are  ${\rm Fr}=10^4$ and $H/d=10$ in both plots. } \label{fig:bet_criRe}
  \end{figure}

Next we show that the phenomenological value $\beta=0.14$ also can be justified by comparing the critical Reynolds number. The experiment performed by Cruickshank \cite{cruick2} yielded the critical value with increasing flow rate, and it corresponds to ${\rm Re}_h^*$ in the present model. As shown in Fig \ref{fig:bet_criRe}(a), ${\rm Re}_h^*$ strongly depends on $\beta$, however $\beta=0.14$ gives ${\rm Re}_h^*=1.38$ which is in accord with the experimental value ${\rm Re}^*_h=1.2$. In contrast to the critical Reynolds number, the dependence of coiling frequency on $\beta$ is relatively slight as shown in Fig \ref{fig:bet_criRe}(b). Thus the discussions in the next section are not strongly influenced by the choice of $\beta$.

\section{Frequency of steady state coiling}\label{sec:steady_freq}
\subsection{Numerical results}
In this subsection, we show coiling frequency as a function of fall height.
The experiments reported by Habibi {\it et al.} \cite{ribe_pre} are performed under ${\rm Re}\sim10^{-5}$, which requires very short time step for computation.
The numerical results shown below are obtained under Re around unity, thus we mention that the comparisons with the experiments are qualitative.

Figure \ref{fig:vis_grav}(a) shows the steady-state coiling frequency as a function of fall height under weak gravity conditions (${\rm Fr}\gg 1$). The frequency decreases with slope $-1$ for $H/d<4$, and becomes almost constant around $H/d=10$. For higher fall height, the influence of gravity becomes significant, and the frequency shows a linear increase. The inset shows the same plot as obtained experimentally, which is in good agreement with the present model.

Next we show the coiling frequency under the strong gravity condition $({\rm Fr\ll 1})$ in Fig. \ref{fig:vis_grav}(b). Here, we choose a characteristic time scale $\sqrt{d/g}$. In this case, the viscous coiling regime is negligible, and the frequency strongly depends on Fr. The frequency-height curve has the slope of $2.0$ while the experiment shows the slope of $2.5$. A clear discrepancy appears for lower fall height, at which the experiment shows a marked discontinuous jump. The present model does not reproduce this behavior. 
\subsection{scaling laws}
\begin{figure}
  \centerline{\includegraphics[width=12cm, keepaspectratio]{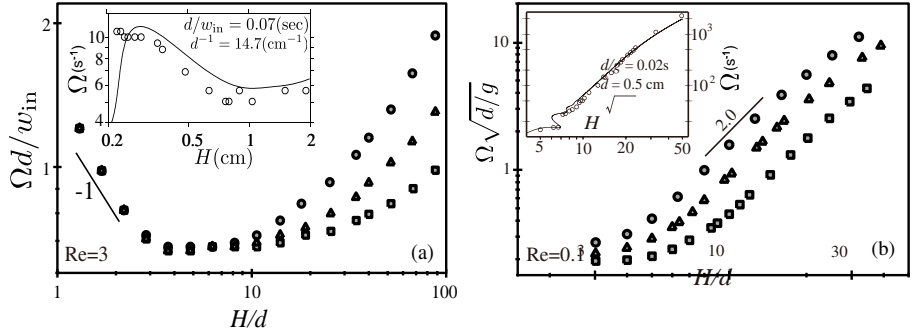}}
  \caption{Coiling frequency as a function of fall height. Insets is the experimental result by Maleki {\it et al.}\cite{ribe_prl}. (a) Coiling under weak gravity with Re$=3$, and Fr$=100$ (circles), Fr=$200$ (triangles) and Fr=$400$ (squares). (b) Coiling under strong gravity with Re$=0.1$ and Fr$=0.01$ (circles), Fr=$0.02$ (triangles) and Fr=$0.05$ (squares).}
 \label{fig:vis_grav}  
 \end{figure}
To compare frequency-height relationships with the the scaling laws [Eqs.(\ref{eq:inertial}-\ref{eq:viscous})], we must estimate the rope radius $a$, which depends on other flow conditions ($Q, g, \nu$ and $H$). Let $\zeta$ be the height at which the flow velocity $w$ reaches a maximum. Under the action of gravity, the internal stress of the fluid rope changes from tensile to compressive at this point, and buckling must occur at $z<\zeta$. Therefore, we reasonably assume $a$ to be the radius at $z=\zeta$ and divide the fluid rope into the ``tail region" ($z>\zeta$) and the  ``buckling region" ($z<\zeta$). 

Radius $a$ is governed by the gravity-induced thinning which is determined by 
the steady-state form of Eq. (\ref{eq:eom_w}):
\begin{equation}
ww'=3\nu w\left(\frac{w'}{w}\right)'-g,\label{eq:thinning}
\end{equation}
To solve this one-dimensional problem, we consider only the flow in the tail region and set $z=\zeta$ as the origin of the $z$ axis. The boundary conditions are $w(H)=-1$ and $w'(0)=0.$ First, we consider the case of weak gravity. For $g=0$, Eq. (\ref{eq:thinning}) has an obvious solution, $w(z)=-w_{\rm in}$. Thus, we seek a solution with the form 
\begin{equation}
w(z)=-w_{\rm in}+g\phi(z).\label{eq:thinning_pt}
\end{equation}
Substitution of Eq. (\ref{eq:thinning_pt}) into Eq. (\ref{eq:thinning}) yields
\begin{equation}
3\nu\phi''(z)+w_{\rm in}\phi'(z)-1=0.
\end{equation}
Solving this equation with the boundary condition $\phi'(0)=\phi(H)=0$, we obtain the perturbation solution 
\begin{equation}
\frac{w(z)}{w_{\rm in}}=-1+\frac{1}{\rm Fr}\left(\frac{z-H}{d}
+{e^{-z/d}-e^{-H/d}}\right),\label{eq:thinning_sol}
\end{equation}
where we set $3/{\rm Re}=1$ for simplicity. The conservation of volume flux gives the radius $a$ as
\begin{equation}
a=d\sqrt{\frac{w(H)}{w(0)}}
=d\left\{1+\frac{1}{\rm Fr}\left(\frac{H}{d}+{e^{-H/d}-1}\right)\right\}^{-1/2} \label{eq:a_ratio}
\end{equation}
Assuming $H/d\ll {\rm Fr}$, Eq. (\ref{eq:a_ratio}) implies that $a\approx d$. Under this condition, the scaling law of viscous coiling may take place, therefore, we obtain $\Omega_V\propto Qd^{-1}H^{-1}$. As the fall height increases, the gravity becomes significant for the tail region. For $H/d\approx {\rm Fr}\gg 1$, Eq. (\ref{eq:a_ratio}) can be approximately written as $a=d(1+gH/{w_{\rm in}}^2)^{-1/2}$.
Substituting this relation into the scaling law (\ref{eq:gravitational}), we obtain 
\begin{equation}
\Omega_G\propto  \left(\frac{gQ^3}{\nu d^8}\right)^{1/4}\left(1+\frac{gH}{w_{\rm in}}\right).
\end{equation}
This indicates that the frequency increase proportional to $H$, which are in accordance with the numerical result shown in Fig. \ref{fig:vis_grav}(a). Thus, the frequency that changes from decrease to increase corresponds to the transition from viscous to gravitational coiling.

\begin{figure}
  \centerline{\includegraphics[width= 7cm, keepaspectratio]{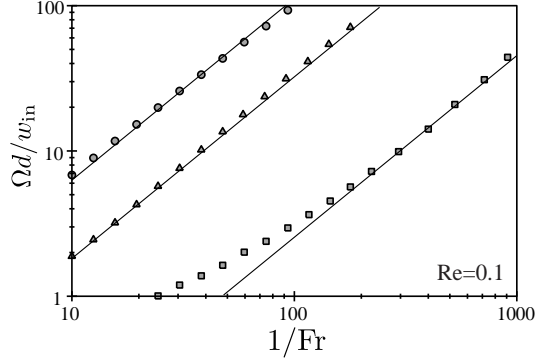}}
  \caption{Frequency versus $1/{\rm Fr}$ for ${\rm Re}=0.1$ and $H=30$ (circles), $H=15$ (triangles) and $H=4$ (squares). Each solid line has the same slope of $1.25$.}\label{fig:omega_g}
\end{figure}

Next we consider that both $\rm Re$ and $\rm Fr$ are much smaller than unity. In this case, the inertia term in Eq.(\ref{eq:thinning}) is negligible, thus $w(z)$ is determined by the balance of gravitational and viscous forces. For the higher fall height, the flow is strongly stretched ($a\ll 1$). In this limit, the force balance implies $a\simeq (Q\nu/g)^{1/2}H^{-1}$ \cite{ribe_prl}. In the buckling region, both gravitational and inertial forces are important.
Eqs. (\ref{eq:inertial}) and (\ref{eq:gravitational}) with the strong stretching condition yield
\begin{equation}
\Omega_G\propto H^2\left(\frac{g^5}{Q\nu^5}\right)^{1/4},~~~
\Omega_I\propto H^{10/3}\left(\frac{g^5}{Q\nu^6}\right)^{1/3}. \label{eq:ss}
\end{equation}
Because the $H$ has similar exponents in these two relations, the gravitational and inertial coiling may behave similarly with changing fall height. Experimentally, $\Omega\propto H^2$ was reported by Cruickshank and Munson\cite{cruick1}. Maleki {\it et al.} also reported $\Omega\propto H^{2.5}$\cite{ribe_prl}.  
To judge whether these results obey the scaling law of gravitational or inertial coiling, we examine the gravity dependence of frequency. In Fig \ref{fig:omega_g}, we show frequency as a function of $1/{\rm Fr}$ for ${\rm Re}=0.1$ and $\rm Fr>10$. Each plot lies on the slope $5/4$. We therefore conclude that the frequency shown in Fig. \ref{fig:vis_grav}(a) corresponds to gravitational coiling. 

The present model does not reproduce the scaling law of inertial coiling, or the discontinuous jump of frequency observed in the gravitational-inertial transitional regime. A possible root of this discrepancy is the boundary condition at the bottom of the fluid rope. In the experiment for the high-frequency coiling regime, the rope rapidly piles up to form a column and it collapses when the height exceeds a critical value \cite{ribe_pre}. In this case, the top of the coiling portion is not stationary, and relation (\ref{eq:bc_bottom2}) may fail. Therefore, the boundary condition at $z=0$ must be improved to render the present model applicable to the inertial coiling regime.

\subsection{Viscous-gravitational transition}
\begin{figure}
  \centerline{\includegraphics[width= 10cm, keepaspectratio]{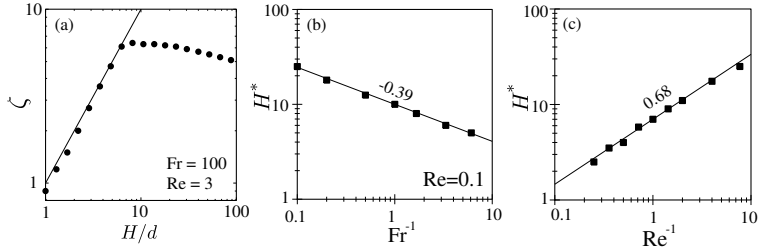}}
  \caption{(a) Numerically calculated value of $\zeta$ for Re$=3.0$ and Fr$=100$.
(b,c) Viscous-gravitational transition height $H^*$ as a function of $1/{\rm Fr}$ and $1/{Re}$, respectively.}\label{fig:bp}
\end{figure}
Within the viscous coiling regime, the coiling frequency decreases with fall hight, while the frequency increases in the gravitational coiling regime. We  discuss the critical fall height $H^*$ at which the frequency is minimized, and which corresponds to the transition between the viscous and gravitational regime. In figure \ref{fig:vis_grav}(a), the transition occurs at $H^*\simeq6d$. However, $H^*$ depends on viscosity and gravitational acceleration. In Fig. \ref{fig:bp}(a), we plot $\zeta$ as a function of $H/d$ for the weak gravity condition ($\rm Re=3$ and $\rm Fr=10^2$). For  shorter fall height, the maximum $w$ appears at the point of injection (plots lie on the line $\zeta=H/d$). Comparing $\zeta$ with the frequency $\Omega d/w_{\rm in}$ in Fig. \ref{fig:vis_grav}(a), we find that the frequency decreases with $H$ only when the relation $\zeta=H/d$ is satisfied. Therefore, viscous coiling appears only when the rope is wholly compressed. Because this feature can be seen for a wide range of Re and Fr as long as the viscous coiling regime exists, we claim that the $H^*$ is identical to the maximum of $\zeta$ in Fig. \ref{fig:bp}(a). The maximum of $\zeta$ can be realized as a relaxation length of which the effect of boundary condition at $z=0$ can travel through a rope. We can uniquely construct a dimension of length using $g$ and $\nu$, as $g^{-1/3}\nu^{2/3}$. 
Therefore,
\begin{equation}
H^*=\max(\zeta)\propto g^{-1/3}\nu^{2/3}. \label{eq:scale}
\end{equation}
We show $H^*$ as a function of $1/{\rm Fr}$ (Fig. \ref{fig:bp}(b)), and $1/{\rm Re}$ (Fig. \ref{fig:bp}(c)). Each plot respectively lies on the slope -0.39 and 0.68 which agrees with the Eq. (\ref{eq:scale}).

\section{Coiling and meandering of dragged fluid rope}\label{sec:meandering}

In this section, we apply the present model to a fluid thread falling onto a horizontally translating surface with speed $U_0$. \cite{webstar} experimentally found that the thread of fluid deposited on the moving surface shows a rich variety of meandering patterns due to the buckling of the rope. 
If the speed of the surface is sufficiently high, the falling thread is strongly dragged to form a viscous catenary \cite{mahadevan2}. In this case, a straight line is drawn on the surface. As $U_0$ decreases, the fluid thread starts to oscillate and sinusoidal or other period-doubling curves appear. Finally, the oscillation reverts to the rotational one of the ordinary fluid rope coiling onto a stationary plate, thereby a cycloid pattern appears on the moving surface.

Although the present model does not include a bottom surface, we can introduce a horizontal drag by adding a control parameter to the boundary condition of Eq. (\ref{eq:bc_bottom1}) as
\begin{equation}
u_x'(0)=\gamma.\label{eq:mod_bc_bottom}
\end{equation}
Here, $\gamma$ is the shear rate emerging as a result of the surface motion. 
We mention that the shear rate may vary during the motion. however,  in the present model, we omit that effect and regard $\gamma$ is positive constant.  
Hereafter, we fix ${\rm Re}=0.1$ and ${\rm Fr}=3.2\times10^{-3}$ (strong gravity condition).
  
\begin{figure}
  \centerline{\includegraphics[width= 10cm, keepaspectratio]{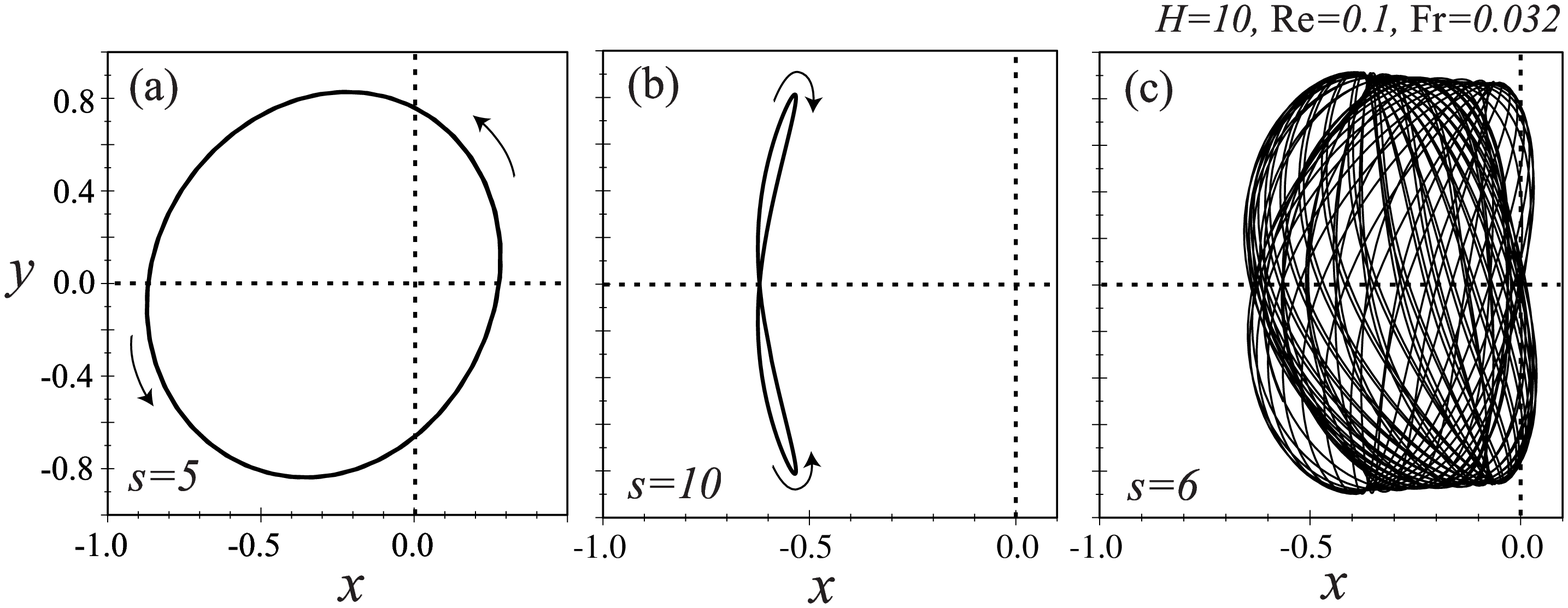}}
  \caption{Steady state trajectories of the bottom of the dragged fluid rope for $H=10$. (a) Trajectory of translated coil state ($s=4$). (b) Trajectories of the meandeing state ($s=10$). (c) A intermediate state between translated coiling and meandering state (s=6).  }
 \label{fig:orbit_meander}
\end{figure}

Changing the value of $\gamma$, we found that the thread exhibits three distinct states: translated coiling (TC), meandering (M) and catenary states.
Figure \ref{fig:orbit_meander} is the trajectory of the bottom of the thread under steady-state oscillation. As $\gamma$ increases from zero, circular coiling is gradually strained [Fig. \ref{fig:orbit_meander}(a)], and subsequently, that the trajectory changes to a figure-eight shape pattern [Fig. \ref{fig:orbit_meander}(b)]. If the shear rate exceeds a critical value $\gamma_c$, the oscillation diminishes in amplitude and the thread forms a stationary catenary (not shown). 

\begin{figure}
  \centerline{\includegraphics[width= 7cm, keepaspectratio]{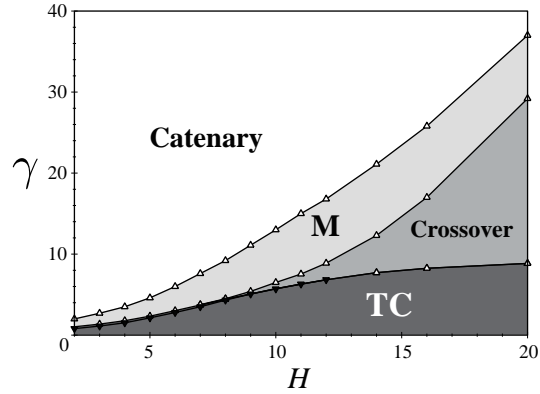}}
  \caption{State diagram in $H$-$\gamma$ plane. Upward (downward) pointing triangles indicate the threshold for increasing (decreasing) shear rate.}
 \label{fig:phase_dia}
\end{figure}

Fig. \ref{fig:phase_dia} is a diagram of the various oscillation modes in the $H$-$\gamma$ plane. The upward (downward) triangles indicate the boundaries of the modes for increasing (decreasing) shear rate. A very weak hysteresis effect can be seen only for lower fall height. The experimentally obtained state diagram is given by \cite{webstar} and more precisely by \cite{morris}.
Here, we note that, because the relationship between $U_0$ and $\gamma$ is unknown, the present model cannot reproduce the patterns of a thread laid down on a moving surface, while the experiments categorize the modes of oscillation on the basis of the patterns. In this study, we compare Fig. \ref{fig:phase_dia} with the state diagram reported by \cite{morris} by assuming that $\gamma$ is proportional to $U_0$, and claim that the trajectories in Figs. \ref{fig:orbit_meander}(a) and (b) respectively correspond to TC and M states.

If the fall height is lower than $H\simeq8$, the transition from TC to M state occurs with discontinuous jump in the amplitude in the $y$ direction. However, for $H>8$, the discontinuity vanishes and a crossover of TC and M states starts to appear. In this region, complex oscillations, tje path of which is not closed, are observed, as shown in Fig. \ref{fig:orbit_meander}(c). It is notable that, in the experiment, M and TC states rarely arise for $H>8.7\pm 0.5$ and complex oscillations named ``stretched coiling" or the ``W state" are observed for higher fall height.

\begin{figure}

\centerline{\includegraphics[width= 6.5cm, keepaspectratio]{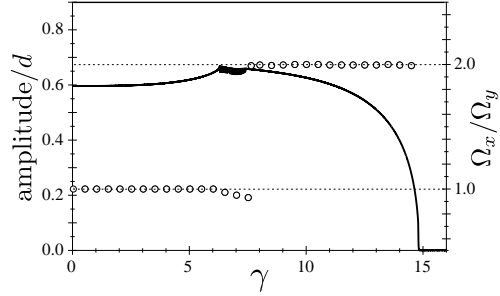}}
\caption{Amplitude in $y$ direction and the ratio of frequency $\Omega_x/\Omega_y$ as a function of increasing $\gamma$. Fall height $H/d=11$.}
 \label{fig:orbit_freq_L11}
\end{figure}  

In Fig. \ref{fig:orbit_freq_L11}, the solid line shows the oscillation amplitude in the $y$ direction with increasing $\gamma$. 
The amplitude gently increases with $\gamma$ in the TC region ($\gamma<6$), and decreases in the M region ($8<\gamma<\gamma_c$). Immediately below $\gamma=\gamma_c$, the amplitude rapidly falls to zero, as  experimentally observed. In the same figure, we also plot the ratio of $x$ to $y$ components of the steady-state frequency $\Omega_x/\Omega_y$. During the transition from TC to M, This ratio jumps from $1/1$ to $2/1$, which agrees with the result of the experiment. We mention that, in the intermediate region ($\gamma\approx 8$), the ratio $\Omega_x/\Omega_y$ slightly deviates from $1/1$.

Experimentally, it is known that $\Omega_y$ increases linearly with $\gamma$ for $\gamma<\gamma_c$, and the frequency at the bifurcation point is in excellent agreement with that of the ordinary coiling state on the stationary surface. However, the present model does not reproduce this tendency; instead, the frequency $\Omega_y$ is a monotonically decreasing function of $\gamma$. 

\section{Summary}\label{sec:summary}
In this paper, we have presented a simple numerical model for the motion of a viscous thread falling onto a plane. We found that the critical Reynolds number $Re^*$ for coil-uncoil transition strongly depends on the boundary condition [Eq. (\ref{eq:bc_bottom2})] at the bottom of the thread. However, the phenomenological parameter $\beta$, which was experimentally determined, gives good estimate of ${\rm Re}^*$. 

The coiling frequency and the scaling laws have been reviewed by classifying the parameter space into areas with weak and strong gravity. In the former, the present model is in accord with both the scaling laws and the results of experiments. However, in the latter case, the inertial regime cannot be reproduced. This may be because, in high-frequency coiling, the boundary condition for $z=0$ is no longer valid. We must overcome this difficulty in our future work. 

The present model can describe the meandering instability of a viscous thread falling onto a moving surface simply by changing the boundary condition at the bottom of the thread. The obtained diagrams of meandering motion and amplitude are in qualitative agreement with the result of experiments.

\end{document}